# Transformation of the singular skeleton in optical-vortex beams diffracted by a rectilinear phase step

Aleksandr Bekshaev[1*], Anna Khoroshun[2], Lidiya Mikhaylovskaya[1]

[1]*Odessa I.I. Mechnikov National University, Dvorianska 2, 65082 Odessa, Ukraine*
[2]*V. Dahl East Ukrainian National University, Pr. Central, 59-A, 93400 Severodonetsk, Ukraine*
[*]*Corresponding author: bekshaev@onu.edu.ua*

Based on the Kirchhoff-Fresnel approximation, we numerically analyze spatial characteristics of the light field formed after a circular Laguerre-Gaussian beam with a single-charged optical vortex (OV) passes the transparent screen with a rectilinear phase step. The main attention is paid to the localization and interactions of the OVs, which form the singular skeleton of the transformed field. The phase-step influence depends on its value and position with respect to the beam axis. Upon "weak perturbation" (low phase step) the main effect is that the OV is shifted from the initial axial position and describes a closed loop when the phase step is monotonously translated across the beam. The "strong perturbation" (the phase step is close to $\pi$) induces topological reactions with emergence and annihilation of additional singularities in the near-axial region of the diffracted beam cross section. These features are interpreted based on the 3D OV trajectories that show an intricate behavior with kinks and "retrograde" segments. The details of the OV migration and singular skeleton transformations reveal the fundamental helical nature and transverse energy circulation in the OV beams. The numerical results obtained in this paper show possibilities for the purposeful control of the singular skeleton characteristics within the transformed beam, and can be useful for the OV diagnostics, OV metrology and micromanipulation techniques.

## 1. Introduction

The study of structured optical fields, in particular, those with optical vortices (OV), is one of the prospective lines of modern optics [1–4]. In scalar paraxial beams, an OV is formed near an isolated point (phase singularity, OV core) with zero amplitude and indeterminate phase of the optical field. The field behavior in the nearest vicinity of the phase singularity is rather standard: the equal-amplitude lines are ellipses centered at the OV core, and upon a round trip near this point, the field phase changes by $2\pi m$ (the integer $m$ is the OV topological charge). The phase singularity is associated with the screw wavefront dislocation and is a center of the transverse energy circulation being the source of the orbital angular momentum [1–5]. The set of such singularities determine the singular skeleton of the beam that qualitatively characterizes the optical field topology, which is relatively stable with respect to external perturbations. The exclusive nature of the singular points makes them well-identifiable, precisely localizable and sensitive "fingerprints" reflecting the optical field state and prehistory [6–9], which is used in numerous metrological applications [10–13]. The beams with OVs find many research and technology utilizations; particularly, they can be employed for the controllable trapping and manipulation of micro-objects (see, e.g. [14–17]) as well as for the information encoding and processing [18,19].

For all fundamental and applied aims, the controllable formation of the OV beam singular skeleton with necessary properties is an insistent problem. It is usually solved by means of purposeful transformations performed to the standard OV beams (e.g., Laguerre-Gaussian (LG) modes [2–4]) with the help of special adjustable optical elements, which can be realized either "physically" or, very efficiently, by means of the computer-driven spatial light modulators (SLM). But prior to employ the unique SLM abilities, one must know how the modeled elements act on the standard input beam, and which modifications of the element's parameters are desirable for obtaining the necessary singular structure at the output. In this context, the detailed studies of the OV beams' transformations performed by various optical units are conducted with high intensity [20–32].

It is known that the efficient control of the singular skeleton can be realized in the processes of the OV beam diffraction [21,22,27–31]. Studies of the edge diffraction of circular OV beams have demonstrated the intriguing dynamics of the phase singularities in the diffracted fields. The multicharged incident OVs (with $|m| > 1$) are decomposed into a set of $|m|$ single-charged "secondary" ones; the secondary OVs evolve along intricate spiral-like trajectories; sometimes the topological reactions occur with generation of "new" OVs and annihilation of "old" ones – in a whole, the singular skeleton of the diffracted beam forms a well developed and controllable set of OVs with the potential of being fruitfully applied in sensitive metrology and optical-trapping techniques.

However, the edge-diffraction schemes are coupled with certain limitations of the available singular skeleton transformations. It is known that the use of other diffractive elements, especially those of the pure phase nature (where the amplitude transmission is homogeneous, and only the phase of the incident beam is modulated) supply many new and potentially utilizable features [20,32–36]. In this context, the simplest phase diffractive elements with rectilinear phase step attract the especial attention. They carry distinct similarities with the well studied edge-diffraction case, and their investigation can be based on the firmly established and reliable footing. Importantly, the corresponding transformations can be easily realized by means of the usual optical elements as well as via the SLM technique. In the current literature, various aspects of the OV beam diffraction by the phase-step transparencies were analyzed [13,22,37–39], and they show promising results, first of all, for the detection of microscopic phase inhomogeneities. In particular, special transformations performed to circular OVs via the phase-step diffraction were fruitfully employed for the sensitive detection of surface micro-structures, precise surface topography characterization and the surface quality control as well as for the interferometric OV diagnostics [13,38,39]. However, possible applications in creation of optical fields with the prescribed and controllable singular structure require systematic investigations of the OV-beam transformations induced by the phase-step diffraction elements. To the best of our knowledge, such studies have not been performed so far.

That is why in this paper we undertake such an attempt. As a generic example of the input OV beam, we consider a Laguerre-Gaussian mode $LG_{pm}$ with zero radial index $p$ [2–4]. This is a usual simplification rather typical for the studies of the OV beams' transformations [23–31,34,36]. But if the incident LG beam is multicharged ($|m| > 1$), the multiple "secondary" OVs separately evolve in the diffracted beam. As a result, specific details of their behavior, although interesting and informative [29–31,38], generally, "mask" the main physical features of the diffracted field associated with the screw wavefront dislocation and the circular energy flow in the incident OV beam. Therefore, in this study we make an additional simplification and restrict ourselves to the case of single-charged incident beam ($|m| = 1$): this not only facilitates the analysis and calculations but promotes to unveiling the generic features of the OV transformations by a phase-step screen.

In the subsequent sections, our main interest concerns the OV displacements, induced by the phase-step transformation. Generally, in the propagating diffracted beam any OV evolves along a certain 3D trajectory (vortex line, or vortex thread) [2,4], and its evolution can be studied in two aspects. First, we can study the 3D trajectory in the whole space behind the screen (so called "$z$-dependent evolution" [31]). This provides a full physical pattern of the OV behaviour but it can only be traced for a limited set of "initial conditions", e.g., phase-step positions with respect to the incident beam axis. In usual experimental approaches [9–17,28,29], the diffracted beam profile in a fixed cross section is registered and analyzed. In these cases, the main interest is associated with the OV migration over the diffracted beam cross section when the phase-step transparency moves across the incident OV beam ("$a$-dependent evolution" [31]). Such situations are considered in detail in sections 3.1 – 3.3. Depending on the phase-step "height" and the propagation distance behind the screen, the cases of "weak" and "strong" perturbation are distinguished. In the latter case, the topological reactions and formation of fully developed singular-skeleton structure are described and discussed. The main regularities of the OV migration in different diffracted beam cross sections, including the far field (Fraunhofer diffraction), are examined.

Additionally, examples of the $z$-dependent evolution are considered in section 3.4 which confirm and generalize the conclusions of previous sections. The final results of the paper open new possibilities for the singular skeleton regulation and show additional ways for the physical characterization and diagnostics of the OV beams. Their potential impact and possible applications are briefly outlined in the Conclusion. The whole presentation is based on the numerical analysis but is confirmed by the asymptotic analytical description of the OV migration in the diffracted beam cross section presented in Appendix.

**2. Description of the phase-step diffraction model**

The general scheme of the incident beam transformation is presented in figure 1 [28–31]. The incident monochromatic paraxial beam propagates along axis $z$, and the spatial distribution of its electric field is described by the function $\text{Re}\left[u(x,y,z)\exp(ikz-i\omega t)\right]$ where $\omega = ck$, $c$ is the velocity of light, $k$ is the wavevector related with the wavelength $\lambda$ as $k = 2\pi/\lambda$, and $u(x,y,z)$ is the slowly varying complex amplitude [3,5]. In the plane $z = 0$, the amplitude-phase corrector (screen) is situated with variable and, generally, adjustable transparency; in further references we call this plane "screen plane", or "initial plane". To highlight this plane, the transverse Cartesian frame at $z = 0$ is furnished with special notations $(x_a, y_a)$, and the incident complex amplitude is denoted as $u_a(x_a, y_a)$; then, the transformation performed by the screen is described by the equation

$$u_a(x_a, y_a) \to T(x_a, y_a) u_a(x_a, y_a). \tag{1}$$

In this paper, we consider the corrector consisting of two parts with different transmittances, divided by the line $x_a = a$ (see figure 1):

$$T(x_a, y_a) = \begin{cases} A, & x_a < a; \\ B, & x_a \geq a. \end{cases} \tag{2}$$

In particular, (2) describes the usual edge diffraction if one part of the screen is opaque (see, e.g., Refs. [26–31] where the situation corresponds to $A = 1$, $B = 0$). Now we analyze the problem in which the screen is homogeneously transparent, $|T(x_a, y_a)| = 1$, but contains a phase step: for $x < a$ the transmission is still $T(x_a, y_a) = 1$ while for $x > a$, a certain coordinate-independent phase $\varphi$ is added to the light field:

$$A = 1, \quad B = \exp(i\varphi) \tag{3}$$

(for convenience, from now on $\varphi$ will be called "height" of the phase step). In figure 1, the phase step in the form of two joint glass plates of different thicknesses is shown for illustrative purposes; in reality, the phase step can be realized by any other means, e.g., with the SLM or a liquid crystal cell.

The complex amplitude $u(x,y,z,a)$ of the propagating diffracted beam in the observation plane positioned at a distance $z$ behind the screen with the phase step located at $x_a = a$ can be found via the Kirchhoff-Fresnel integral [29,31,40,41]

$$u(x,y,z,a) = \frac{k}{2\pi i z} \int_{-\infty}^{\infty} dy_a \int_{-\infty}^{\infty} dx_a T(x_a, y_a) u_a(x_a, y_a) \exp\left\{\frac{ik}{2z}\left[(x-x_a)^2 + (y-y_a)^2\right]\right\}$$

$$= \frac{k}{2\pi i z} \int_{-\infty}^{\infty} dy_a \int_{-\infty}^{a} dx_a u_a(x_a, y_a) \exp\left\{\frac{ik}{2z}\left[(x-x_a)^2 + (y-y_a)^2\right]\right\}$$

$$+ \frac{k}{2\pi i z} \exp(i\varphi) \int_{-\infty}^{\infty} dy_a \int_{a}^{\infty} dx_a u_a(x_a, y_a) \exp\left\{\frac{ik}{2z}\left[(x-x_a)^2 + (y-y_a)^2\right]\right\}. \qquad (4)$$

Our task is to study the optical field formed behind the transparency (2), (3) when the incident field is a circular beam with OV. As was explained in the Introduction, a generic example of such beams is provided by the $LG_{0m}$-beam with zero radial index and, to avoid unnecessary complications and to unveil the main physical features of the OV-beam diffraction associated with its helical nature, we consider the single-charged incident beam. In this case, the integral (4) can be reduced to explicit analytical expressions (see, e.g., [23,24,26]) but its numerical evaluation is also appropriate; anyway, the exact $(x, y)$ coordinates of the OV cores are found numerically as the zeros of amplitude – solutions to the equation $u(x,y,z) = 0$ [28–31].

According to the experimental scheme of [28,29] (see figure 1), we consider the $LG_{0m}$ incident beam with $m = -1$ (clockwise energy circulation as seen from the positive end of the $z$-axis, see figure 1). Additionally we suppose that the screen plane coincides with the beam's waist plane $z = 0$, that is

$$u_a(x_a, y_a) \propto \frac{x_a - iy_a}{b^2} \exp\left(\frac{x_a^2 + y_a^2}{2b^2}\right) \qquad (5)$$

where $b$ is the Gaussian envelope waist radius [2,5]. Note that if the incident complex amplitude is modified according to the equation

$$u_a(x_a, y_a) \rightarrow u_a(x_a, y_a) \exp\left(ik\frac{x_a^2 + y_a^2}{2R}\right) \qquad (6)$$

(e.g., the plane wavefront is replaced by the spherical one preserving the same intensity profile), the diffracted beam pattern can, by means of the simple scaling transformation, be obtained from that derived for the initial complex amplitude $u(x_a, y_a)$ [26,30]. Therefore, despite the accepted plane-front limitation, results of the present analysis can be easily adapted to arbitrary converging or diverging beams with non-zero wavefront curvature $R^{-1}$ in (6).

Other useful generalizations follow from the symmetry of the circular OV beams and the geometric symmetries inherent in the situation presented in figure 1. For example, the inversion of the incident OV sign ($m \rightarrow -m$) means that in the integrands of (4) $u_a(x_a, y_a) \rightarrow u_a(x_a, -y_a)$, which results in the replacement $u(x, y, z, a) \rightarrow u(x, -y, z, a)$ for the diffracted field. Therefore, for incident beams with positive OVs, all the patterns presented below in figures 2 – 8 should be mirror-reflected with respect to the $x$-axis. A bit longer chain of arguments including the LG beams' property $u_a(-x_a, -y_a) = (-1)^m u_a(x_a, y_a)$ and proper inversions of signs of the integration variables in (4) leads to a conclusion that for the opposite phase step, $\varphi \to -\varphi$, the output beam complex amplitude can be obtained from the result of (4) through the transformation

$$u(x,y,z,a) \to e^{-i\varphi}(-1)^m u(-x,-y,z,-a). \tag{7}$$

This permits us to consider only the case of positive $\varphi$. And, of course, as phase shifts $-\varphi$ and $2\pi - \varphi$ are physically equivalent, the range $0 < \varphi \leq \pi$ enables to describe all possible phase-step-induced transformations.

For the incident beam (5), the diffraction integral (4) can be expressed via the set of dimensionless spatial parameters where all the transverse sizes (including the step position $a$) are expressed in units of the beam waist radius $b$ (or its derivatives as given by equation (10) below) and all the longitudinal parameters – in units of the Rayleigh range [2,3,29]

$$z_R = kb^2. \tag{8}$$

This scaling scheme is systematically used in further presentation and exposition of the results. Simultaneously, in all numerical calculations we supposed the specific conditions of Refs. [28,29] where the incident $LG_{0,-1}$ beam of the He-Ne laser radiation with parameters

$$k = 10^5 \text{ cm}^{-1}, \quad b_0 = 0.232 \text{ mm}, \quad z_R = 53.8 \text{ cm} \tag{9}$$

was employed. In some examples, in order to directly address the real experimental situation, the absolute values of the coordinates are also presented, which can always be associated with the dimensionless data by means of equations (8) and (9).

## 3. Characteristics of the transformed beam and singular skeleton structures

### 3.1. General features of the diffracted beam profile

In this section we analyze the influence of the phase-step transparency (figure 1) on the spatial characteristics of the diffracted beam propagating behind the screen. The general illustration of the corresponding beam transformations is given in figure 2 where the diffracted beams' intensity and phase profiles are presented. The phase profiles in the right column of figure 2 are expressed by the equiphase lines of different colors: the local phase grows from violet to blue, from blue to cyan, and so on, with increment 1 rad. The complicated multi-sheet phase surfaces of the OV beams cannot be projected onto a plane without cuts which are seen as tight "bundles" of differently colored lines; ends of the cuts are just the points of phase singularities where the field amplitude is zero, i.e. the OV cores (denoted in figure 2 by $V_j$ with $j = 1$ to 4).

1st row of figure 2 expectedly shows that the phase step $\varphi = \pi/3$ makes a weaker perturbation of the beam than the opaque-screen edge in the case of edge diffraction [20–31]: even at $a = 0$ (the phase edge crosses the beam center), the diffracted beam preserves the ring-like shape, although deformed; the main visible influence of the phase step is that the OV $V_1$ is shifted from the initial axial position. Case of $\varphi = \pi$ (2nd and 3rd rows of figure 2) show much higher perturbation: even when the step is positioned rather far from the center ($a = 1$ in units of $b$ for the bottom row), the beam structure is essentially modified, which is well seen in the intensity patterns of figures 2c and 2e. Not only the axial OV is displaced ($V_1$ in figures 2d and 2f) but also additional OVs appear. Generally, this is not surprising: in any case of the sharp-edge diffraction, "occasional" OVs emerge at the diffracted beam periphery simultaneously with the diffraction fringes [2,21,22,26,27] (normally, with further post-screen propagation these OVs rapidly migrate to the periphery of the beam cross section, and they are not considered in our present context). But now, in figure 2, together with the presumably "occasional" peripheral OVs ($V_3$ and $V_4$ in figure 2d), new near-axial singularities appear ($V_2$ and $V'$ in figures 2d, f): the complex singular skeleton is formed.

Figures 2c and 2d illustrate another important feature of the charge-1 LG beam diffraction by the $\pi$-step. In case of $a = 0$, just behind the screen, the edge wavefront dislocation induced by the phase step crosses the center of the screw dislocation inherent in the OV, and both singularities partly compensate each other. The connection between different sheets of the helical wavefront is destroyed, and the wavefront becomes although a discontinuous but simply connected surface (see the inset in figure 2c): upon any round-trip near the beam axis, the beam phase returns to its initial value, which corresponds to the effective topological charge $m = 0$ of the diffracted beam. This property conserves during the free propagation, and that is why the diffracted beam carries equal numbers of "positive" and "negative" OVs (see figure 2d). Notably, the situation of $\varphi = \pi$ and $a = 0$ is exceptional, and any violation of either of these conditions destroys the topological charge compensation. For example, the inset in figure 2a shows that in case of $\varphi = 2\pi/3$ and $a = 0$, the wavefront behind the screen retains its helical multi-sheet nature, and the phase still changes by $2\pi$ after a round trip near the axis. Accordingly, the "sum" topological charge of the diffracted beam is the same as of the incident beam (see, e.g., figures 2a, b). However, the $\varphi$-step phase singularity is also coupled with the zero-amplitude line immediately behind the screen, which is similar to the zero-amplitude line in the plane of formation of a fractional OV (cf., for example, [42,43]). In contrast to the fractional OV case, no chains of OVs are formed in the course of further diffracted beam propagation; the zero-line is merely blurred out but its rudiment can be seen at moderate distances behind the screen (see the relative intensity minimum of the bright ring in the lower half of figure 2a).

Figure 2 shows that the whole pattern of the diffracted beam perturbed by the rectilinear phase step is rather intricate, and its thorough study is a special task. In this paper we will concentrate on the main singular skeleton characteristics: positions of the near-axial OVs, directly associated with the OV carried by the incident beam, and their migration over the beam cross section when the phase step moves across the incident beam transverse profile.

In the calculations, we investigate the situations when the phase-step screen moves from $a = 3$ towards $a = -3$ (theoretically, from the "right infinity" $\infty$ to the "left infinity" $-\infty$ when seeing against the beam propagation). The figures 3 – 6 below show the migration of OVs over the beam cross section in the chosen observation planes situated at $z = 0.56z_R$ (30 cm for the incident beam with parameters (9)), $1.5z_R$ (80 cm), $3.7z_R$ (200 cm, in figure 6) and in the far field (Fraunhofer diffraction [40,41]). For convenience and in compliance with the previously used terminology [13,28,29,31], the maps of OV migration are frequently called "trajectories"; this will cause no confusion with the "true" 3D trajectories discussed in section 3.4. To make the data relating to different propagation distances more comparable, the transverse coordinates are expressed in units of the current Gaussian envelope radius (see (5))

$$b_c = b\sqrt{1+\left(\frac{z}{z_R}\right)^2} . \qquad (10)$$

For the far-field patterns, this normalization means that the angular far-field coordinates are expressed in units of the incident beam divergence $\gamma = (kb)^{-1}$: $(x/b_c, y/b_c) \rightarrow (x/z)kb, (y/z)kb$.

### 3.2. OV migration in the case of weak perturbation

In further presentation, it is convenient to classify the phase-step influences on the OV positions by two categories. In the first case, the OV core shifts from the nominal axial position and evolves along a certain smooth curve. We will refer to such situations as to the "weak perturbation" pattern, and some of them are presented in this section. In the opposing case, the diffracted beam structure is modified stronger so that the OV trajectory becomes discontinuous, the events of OV birth, annihilation and other topological reactions take an essential place; this will be termed "strong perturbation". Of course, this division involves not only the transparency shaping; the "strength" of the perturbation depends on the diffracted beam propagation, and sometimes the same transparency performs a strong perturbation just behind the screen ($z < z_R$) but the weak one in the far field ($z \gg z_R$). But, generally, the beams passed phase steps with $\varphi < 0.7\pi$ can be considered as weakly perturbed, and the typical situations are illustrated by figure 3. These OV trajectories qualitatively differ from the cases of higher $\varphi$ that will be discussed further.

In figure 3 we see that initially, while $\infty > a > 2$, the only consequence of the phase-step perturbation is that the axial OV in the diffracted beam is slightly displaced from the nominal position (at the $z$-axis) and moves along the spiral trajectory with theoretically infinite number of coils. Note that when the phase step advances towards the axis, the spiral of the OV trajectory evolves oppositely to the energy circulation in the incident beam (cf. the arrow in figure 1). This spiral-like evolution is a consequence of the interference between the unperturbed incident beam and the edge wave generated by the sharp edge [41] (here – by the rectilinear phase step) and is similar to what was observed in the usual edge-diffracted beams [29,31]; it is well described by the asymptotic analytical model that can be derived for the case $a \gg 1$ in units of $b$ (see Appendix). In this part of the trajectory, corresponding to $\infty > a > 2$, the absolute OV displacement inversely depends on $z$ [29–31]; that is why the segment of spiral-like OV evolution at $a > 2$ can be traced in figure 3a but becomes less perceptible in figure 3b and completely disappears in the far-field situation of figure 3c. Anyway, in such segments, the OV displacements are small and of a minor practical interest, at least for the simple situation of the single-charged LG beam considered here.

With a subsequent phase-step translation, the OV deviates from the spiral trajectory (approximately when $a = 1$ in figure 3a) and moves rather far from the axis "escaping" from the approaching phase step. Expectedly, the absolute value of this deviation is higher for the higher phase steps (stronger incident beam perturbation for the same step positions). This sort of evolution continues until $a = 0$ (the phase step crosses the incident beam center), after which the OV motion is slowed down and turns back toward the axis. The "return point" occurs approximately at $a = -0.4$ for $z = 0.56 z_R$, $a = -0.3$ for $z = 1.5 z_R$ and exactly at $a = 0$ in the far field.

During the following screen translation towards $a = -\infty$, the OV trajectories make closed loops, ultimately returning to the nominal beam axis. It is seen that the "global" OV circulation along the whole loop is always clockwise, i.e. matches the energy circulation in the incident beam (according to relation (7), for negative $\varphi$, the same screen translation would induce a counter-clockwise OV migration). The size of the loop is higher for higher $\varphi$ (cf. the green, blue and red curves in figure 3). With growing phase step, the diffracted beam perturbation becomes sufficient for inspiration of the trajectory discontinuities and associated topological reactions: in case of $\varphi = 2\pi/3$ and at small distance behind the screen (red curve in figure 3a), the OV trajectory experiences a "jump" similar to those investigated in Refs. [29–31] (in fact, this situation is "transient" to the strong-perturbation case considered below, cf. the jumps of the red and green trajectories in figure 6).

Actually, when the phase step approaches the position $a = -0.96$ (point F in figure 3a), an OV dipole emerges in the remote point E of the diffracted beam cross section. Then, the oppositely charged (positive) dipole member rapidly moves against the "main" OV evolution (black curve arrow in figure 3a) and annihilates with the "initial" OV in the point D marked by the asterisk (approximately when $a = -0.99$). (This oppositely-charged OV can be called "virtual" [31]: it is "short-living" (exists within a limited range of the phase-step positions $a$), and its main role is to assist the interaction "transferring" the "main" OV from F to E, which resembles the virtual particles in quantum physics). Simultaneously, the negatively charged dipole member slowly evolves clockwise from the point E thus forming the trajectory continuation. These processes are coupled with the rapid transformations of highly anisotropic OVs and are thus difficult for the detailed examination numerically as well as in experiments; for this reason, they were conventionally united into a single event named "jump" [31], and the "moment of jump" was defined as the value of $a$ (or $z$, in case of the $z$-dependent evolution, see section 3.4) at which the new pair of OVs is born in point E. Note that in figure 3 the jump is observed only on the red curve and at the closest post-screen distance $z$ because only in this case the jump criterion $|M| > 1$ is realized (see Appendix). In contrast to the case of the edge-diffracted OV beams [29–31], in the red curve of figure 3a the jump occurs at negative $a$ and in the upper half-plane, in full agreement to equations (A14) and (A10).

When $a \to -\infty$, all the trajectories return to the axis, which is not surprising because the screen becomes again a homogeneous transparency and performs no perturbation to the incident LG beam. Remarkably, for high negative values of $a$, all the trajectories demonstrate a behavior quite "symmetrical" to that observed when $a$ decreased within the interval $\infty > a > 2$: the trajectories approach the point ($x = 0$, $y = 0$) making theoretically infinite number of squeezing spiral coils in the interval $-2 > a > -\infty$ (except for the "degenerate" far field situation in figure 3c). Oppositely to the stage when $a$ decreased, now, when the phase step moves away from the incident beam axis, the spiral wraps clockwise, according to the energy circulation (see figure 1). This agrees with the predictions of the asymptotic model (A13), (A14).

Note also the regular change in the shape of the OV trajectories with growing propagation distance: from the "banana-like" loops carrying distinct reminiscence of the spiral evolution near the coordinate origin (figure 3a for $z = 0.56 z_R$) via the "straightened" contours at $z = 1.5 z_R$ (figure 3b) towards the "cigar-like" patterns in the far-filed (figure 3c). As the far field intensity distribution is always symmetric with respect to the $x$-axis normal to the diffraction edge (this is a characteristic feature of the far-field diffraction once the incident beam waist coincides with the screen plane [27,30]), the far-field OV trajectories expectedly obey this symmetry. Likewise, it is not surprising that the maximum absolute displacement of the OV is reached when $a = 0$. It is remarkable, however, that all the far-field trajectories form distinct cusps at $a = 0$ (see figures 3c and 5).

### 3.3. Singular skeleton evolution and topological events caused by the strong phase-step perturbation

When the phase-step value approaches $\varphi = \pi$ (see (2), (3)), the axial OV displacements expectedly become higher but, what is more, the OV trajectory induced by the monotonous screen translation within the range $\infty > a > -\infty$ acquires additional branches and experiences topological reactions. Actually, already the red curve in figure 3a contains some features transient to the strong perturbation behavior which is exhaustively presented in figure 4.

At the initial segment, while $\infty > a > 2$, the axial OV is slightly displaced from the axial position and evolves along the spiral trajectory similarly to what was seen in figure 3. This part of the trajectory is not shown in figure 4a. With further screen translation, while $a < 2$, the OV evolves along the magenta trajectory marked $V_1$ (see figures 4a and 4b). However, when $a$ reaches $a \approx 1$, an additional pair of OVs is born in the 4th quadrant in point B (this situation is illustrated by the dashed ellipse in figure 2f). The OV $V_2$ of the same charge as the initial one moves along the magenta curve while the oppositely charged "accompanying" OV $V'$ (black) moves to the lower beam periphery and eventually disappears (its motion can be numerically traced up to $a \approx 0.3$ but the "far" segment of its trajectory is not shown). So, when the screen position corresponds to $1 > a > 0.4$ (this interval slightly varies with the propagation distance and looks a bit smaller at $z = 1.5 z_R$, see figure 4b), the three OVs are present within the central area of the diffracted beam cross section: $V_1$, $V_2$ and $V'$. With further advance of the screen, $0.4 > a > -0.4$, only two OVs $V_1$, $V_2$, with the topological charges $m = -1$ (equal to the initial one), survive and form the left ($V_1$) and right ($V_2$) trajectory branches. Further, when $a$ approaches $-0.3$, the oppositely charged OV becomes visible at the upper periphery ($V''$) and moves to meet $V_1$ (in figures 4b and 4c, only its part for $a < -0.7$ is shown). $V_1$ and $V''$ annihilate at $a \approx -1$ (point A), and for $a < -1$, again, only one OV exists ($V_2$); with $a$ decreasing further (the phase-step line goes outside the beam cross section), this OV approaches the origin via describing an infinite (theoretically) number of spiral coils, quite similar to the final ($a < -1.5$) segments of the green, blue and red loops in figures 3a and 3b.

The remarkable central symmetry between the trajectories of $V_1$ and $V_2$, $V'$ and $V''$ is explained by the fact that for $\varphi = \pi$ the phase steps $\varphi$ and $-\varphi$ are physically equivalent, and condition (7) leads to the intensity profile symmetry

$$|u(x,y,z,a)|^2 = |u(-x,-y,z,-a)|^2.$$

In this context, the clockwise evolution of $V_1$ at $1 > a > -1$ agrees with the energy circulation handedness and thus follows the general rule noticed for the closed loops observed under the weak-perturbation conditions and $\varphi > 0$ (figure 3). Similarly, the counter-clockwise circulation of $V_2$ in the region $1 > a > -1$ complies with the same rule for $\varphi < 0$ (see section 3.2).

The whole pattern acquires additional mirror symmetry in the far field (figure 4c). The point B moves to the vertical $y$-axis, and birth of the OV dipole $(V_2,V')$ occurs exactly at the moment when the OV $V_1$ approaches this point. During further evolution, $V'$ moves to the far negative-$y$ periphery and leaves the central part of the beam cross section whereas the OVs $V_1$ and $V_2$ describe the mirror-symmetric trajectories (note the cusps at $a = 0$ similar to what was observed in the far-field trajectories of simpler curves in figure 3c; cf. also the far-field trajectories of $V_1$ in figures 5 and 6). At last, approximately at $a = -0.4$, an oppositely charged OV appears at the positive-$y$ periphery $(V'')$, and the annihilation of $V_1$ and $V''$ happens in point A, which now lies on the $y$-axis. Exactly at the moment of annihilation, the vortex $V_2$ passes the point A, abruptly turns downward and finishes at the coordinate origin.

Alternatively, we can consider how the far-field pattern of figure 4c develops when the observation plane is fixed but the phase step parameter $\varphi$ approaches the limit value $\pi$ from below (figure 5). For example, let us focus on the case $\varphi = 2\pi/3$ presented by the red lines. It is seen that while the "main" OV $V_1$ describes the closed trajectory in the $x < 0$ half-plane (the red "cigar" shown in figure 3c is reproduced, in another scale, in figure 5), at the moment when $a \approx 0.7$, in the opposite half-plane an OV dipole emerges (in the "red" point B). The "accompanying" dipole member $V'$ goes to the far periphery but its counterpart $V_2$ regularly evolves along the "gull-wing" trajectory A0B until it vanishes in the annihilation event in the "red" point A (note that this evolution is, generally, opposite to the clockwise energy circulation). This migration of $V_2$ is not very noticeable since at any moment $V_2$ is positioned relatively far from the beam center, in the low-intensity region. At first sight, it looks as one of the multiple OVs emerging in any diffraction pattern formed behind a screen with a sharp inhomogeneity of transmission [26,27] but as we see, the OV $V_2$ survives up to the far field conditions and behaves as a regular "partner" of the "main" one $V_1$.

The association between $V_1$ and $V_2$ becomes more evident if we consider the singular skeleton transformation when $\varphi$ approaches closer to $\pi$. According to figure 5, for $\varphi = 5\pi/6$ the evolution is rather similar to what is observed for $\varphi = 2\pi/3$ but the cyan cigar-like loop in the left half-plane gets larger, the "cyan" points B and A are situated closer to the beam center, and, on the whole, the "cyan" trajectories tend to the symmetric limit case represented by the magenta curves: the cyan loop – from the inside, the "gull-wing" – from the outside. When $\varphi \to \pi$ still closer, the symmetric structure of figure 4c is formed: points B and A take their places on the vertical axis, and the "cigar" and "gull-wing" curves merge with the "left" and "right" magenta contours, correspondingly. But this reasoning leads to another interpretation of the topological reaction occurring in the "magenta" point A: in contrast to the situation considered in figure 4 (when $\varphi = \pi$ was fixed and the observation plane moved to the far-field zone), now it is reasonable to assume that in point A the OV $V_2$ annihilates with the $V''$ whereas the OV $V_1$ turns downward and continues its motion to the beam center forming thus a closed loop, as is observed for other far-field patterns (cyan and red curves in the left half-plane of figure 5).

Such double interpretation is possible because the far-field situation for $\varphi = \pi$ represents a degenerate case which can be realized by two different limit transitions: $z \to \infty$ while $\varphi = \pi$ is fixed, and $\varphi \to \pi$ while $z = \infty$ is fixed (cf. also the discussion in 3rd paragraph of section 3.1). Actually, each of the topological reactions in points B and A involves four OVs on the equal footing. In point B, there is a single input member ("upper" $V_1$) and three output ones: "left" $V_1$, $V_2$ and $V'$; likewise, in point A the three input ones ($V_1$, "right" $V_2$ and $V''$) and one output ("lower" OV that moves from A towards the center along the vertical axis) are present. In figure 4c this lower OV was interpreted as a continuation of $V_2$, which is compatible with the continuous transition from the patterns presented in figures 4a and 4b. In figure 5, oppositely, treating the "lower OV" as a continuation of $V_1$ emphasizes the kinship between the magenta curve and the cyan and blue curves.

Analogs of the OVs marked $V_2$ in figures 4 and 5 exist not only in the far field and not only for $\varphi = \pi$. Generally, presence of the second OV of the same sign as the incident one during a certain part of the transverse phase-step translation is a characteristic feature of the "strong perturbation" conditions. This is illustrated by figure 6 where the phase step $7\pi/8$ is close to the limit value $\varphi = \pi$ and the use of $z$-dependent, normalized by (10), coordinates enables immediately comparing the OV trajectories at different post-screen distances. It is seen that while $V_1$ evolves along the closed trajectories (whose shapes gradually change with growing distance $z$), at a certain, $z$-dependent, stage of the phase-step translation, its "partner" $V_2$ emerges in the peripheral area of the 4$^{th}$ quadrant (points B) and migrates across the right-hand part of the beam cross section until it vanishes due to annihilation. (To avoid the figure overloading, the corresponding oppositely charged dipole members accompanying the birth ($V'$) and annihilation ($V''$), well seen in figures 4 and 5, are omitted in figure 6). The "life times" of the singularities $V_2$ (seen from the numbers near the initial and final points of their trajectories) depend on $z$ and are limited but, generally, the two OVs, $V_1$ and $V_2$ are readily observable when the phase-step is positioned near $a \approx 0$ and are the constitutive components of the diffracted-field singular skeleton.

Qualitatively, the pattern of figures 5, 6 is also consistent for the "weak perturbation". For example, each loop in figures 3a and 3b can be considered as a trajectory of $V_1$ for which the "partner" OV $V_2$ emerges, migrates and vanishes somewhere at the far right-hand periphery where the light intensity is very low (figure 5 shows that the more the phase step $\varphi$ differs from $\pi$, the more is the distance of the corresponding "partner" OV trajectory from the beam center). Practically this means that for $\varphi < 2\pi/3$ only the "main" OV $V_1$ can be well recognized and used for applications while the peripheral "partner" of $V_1$, the vortex $V_2$, is hardly distinguishable from the "occasional" OVs emerging with the diffraction fringes.

Comparison of figures 6 and 4a discloses additional details of the mechanism of the singular skeleton transformation with growing $\varphi$ and confirms the deep intrinsic affinity between the "main" $V_1$ and "partner" $V_2$ singularities. Indeed, let us consider the red curve of figure 6 describing the situation of $z = 0.56 z_R$ and the corresponding figure 4a for $\varphi = \pi$. Obviously, when $\varphi$ will approach still closer to $\pi$, $7\pi/8 < \varphi < \pi$, the jump in the red curve of figure 6 will become still more articulate: the distance between the "red" points D and E grows, and the event of the dipole birth in point E takes place at earlier stages of the screen translation (higher values of $a$).[1] After all, the topological connection between points D and E will be destroyed but the connection between points E and A of the red curves establishes (cf. [31]): $V_1$ annihilates near the point D with the "accompanying" OV approaching from the upper periphery ($V''$ in Fig. 4a) while the "virtual" member of the dipole born in E moves no longer towards F and D (as was in Fig. 3a) but to the "red" point A that, with $\varphi \to \pi$, approaches closer to the beam center. As a result, the "red" point D of figure 6 transforms into the point A of figure 4a, whereas the "red" points E and A of figure 6 merge together and form a continuous trajectory interpreted as $V_2$ in figure 4. Consequently, from a more general point of view, the trajectories of $V_1$ and $V_2$ can be considered as separate branches of the certain "combined" path of the migrating singularity, which belongs to a complex multi-sheet abstract surface [44].

---

[1] In particular, under such conditions the usual interpretation of the $V_1$ evolution between points F and E as a "jump" becomes less appropriate, see figure 8 and the corresponding discussion below.

## 3.4 3D vortex lines and the $z$-dependent singular-skeleton evolution

In previous sections, main attention was focused on the OV migration observed in fixed cross sections of the diffracted beam while the phase step is translated across the incident beam ($a$ changes from $+\infty$ to $-\infty$, see figure 1). Accordingly, the results presented in figures 3 – 6 are rather "instrumental" and address the usual experimental situations or applications for the OV metrology and optical trapping where keeping the observation plane fixed is more convenient [8–17,28,29]. However, the singular skeleton of the diffracted beam actually forms a complex 3D structure, and its investigation is especially helpful for the physical analysis of the diffraction-induced OV-beam transformations [2,4,20,27,30,31]. In this section we consider several examples of the 3D singular skeletons generated at fixed phase-step positions $a$. The phase-step heights ($\varphi = \pi/2$, $\varphi = 2\pi/3$ and $\varphi = 5\pi/6$) are chosen for better illustration of the transition between the "weak" and "strong" perturbation conditions classified in the previous sections.

Figure 7 shows the OV migration in case of $a = 0.8$. The "true" 3D evolution is presented only for $\varphi = 2\pi/3$ in the inset (note that the longitudinal coordinates are expressed in units normalized by (10), due to which the infinite interval (0, $\infty$) is mapped onto the finite range (0, $z_R$)). The trajectories depicted in the main panel are the transverse projections of the "true" 3D trajectories.

At first glance, the evolution of the "central" OV $V_1$ (cf. figures 5, 6) shows no peculiar features. As is typical for other cases of the OV diffraction [28–31] and is dictated by the asymptotic model presented in the Appendix, at the initial stage ($z < 5$ cm), the OVs $V_1$ migrate along spirals evolving oppositely to the energy circulation in the incident beam and making, theoretically, an infinite number of coils near the coordinate origin; these parts of the trajectories are not shown in Fig. 7 because of the very small absolute OV displacements. One can also remark that final points of the trajectories corresponding to $z = \infty$ agree with the $V_1$ positions for $\varphi = 2\pi/3$ and $\varphi = 5\pi/6$ in figure 5.

However, analysis of the 3D pattern helps to disclose the physical nature of the "partner" and "accompanying" vortices $V_2$ and $V'$: actually, these belong to a single continuous "combined" vortex line. The inset shows that the vortex line $V_2$ emerges at low $z$ (most probably, due to the usual mechanism of the OV formation from the speckle pattern of the diffraction fringes [2,21,22,26]; this stage of the diffracted beam evolution takes place in the near-field region $z < 0.1 z_R$ and is not considered here), evolves along the beam propagation up to the plane $z_1$ (transverse semitransparent plane in the inset) and then "turns back" so that the $V'$ trajectory can be treated as a sort of retrograde vortex-line motion with respect to the "overall" beam direction indicated by the $z$-axis. In the moving coordinate frame associated with this combined ($V_2 + V'$) vortex line (in which the longitudinal axis coincides with the tangent straight line), the energy circulation direction is constant [2,4] but in the laboratory frame the retrograde segment looks as if the topological charge of $V'$ is opposite. The similar features of the 3D patterns are typical to other cases of the OV-beam diffraction discussed earlier [20,31,32].

The inset of figure 7 immediately illustrates the case of $\varphi = 2\pi/3$ in which the "critical point" of the 3D trajectory (where the trajectories of $V_2$ and $V'$ converge and the cross-section plane is tangent to the 3D vortex line) occurs at $z = z_1 = 240$ cm $= 4.5 z_R$. For $\varphi = \pi/2$ the pattern is similar but the vertical tangent plane occurs earlier, at $z_1 = 170$ cm $= 3.16 z_R$, while for $\varphi = 5\pi/6$ trajectories $V_2$ and $V'$ never converge and the continuous combined vortex line is not formed ($V_2$ and $V'$ exist separately even in the far field, which is seen in the main panel of figure 7 and in figure 5). This is an additional attribute distinguishing the strongly perturbing phase steps ($\varphi > 0.7\pi$) from the weakly perturbing ones (cf. section 3.2).

In figure 7, which shows the picture observable when the observation plane moves towards $z = \infty$, the convergence of trajectories $V_2$ and $V'$ looks as the OV-dipole annihilation. However, if the observation plane is fixed, the similar event can happen when the phase-step position changes: e.g. while $a$ decreases from $+\infty$ to $a = 0.8$ (the phase step moves from the left beam periphery towards its axis, see figure 1), the combined 3D curve ($V_2 + V'$) becomes more "prolonged", and the critical point moves to higher distances behind the screen ($z_1$ grows, see figure 9a). Therefore, in a fixed observation plane, an event, in which the combined 3D trajectory approaches this plane and crosses it, is naturally interpreted as the OV-dipole birth. For the far-field observation plane, this situation is described by figure 5, and, obviously, the annihilation of $V_2$ and $V'$ for $\varphi = 5\pi/6$ and $\varphi = 2\pi/3$ in figure 7 and the birth of $V_2$ and $V'$ in figure 5 are manifestations of the same physical phenomenon: intersection of the observation plane and the combined 3D vortex line. For this reason, the annihilation points in figure 7 are marked by the same letters "B" as in figure 5.

Figure 8 illustrates another situation that happens with the same phase screens but if the phase step is positioned on the other side with respect to the incident beam symmetry plane parallel to the phase-step line (see figure 1) – at $a = -0.8$ (according to figures 3a and 6, near such phase-step positions the $a$-dependent trajectories in fixed cross sections experience jumps). The red and cyan curves in figures 8a, 8b and 8c show that the nature of the jumps is essentially the same as in case of the edge diffraction [31]. The "central" OV $V_1$ evolves along the 3D vortex line which, at certain conditions, bends backward and makes a Z-like meander so that some transverse planes cross this line in three points instead of the usual one. This Z-like segment is bordered by the critical points where a transverse plane is tangent to the 3D vortex line (longitudinal positions of these critical points are highlighted by the semitransparent transverse planes $z_1$ and $z_2$ in figures 8b and 8c). Note that at the weak-perturbation conditions ($\varphi = \pi/2$ and $a = -0.8$, blue curve) such a Z-like segment does not occur, and is thus not shown in the corresponding 3D picture (more correctly, it can only exist at lower $|a|$ and smaller propagation distances, see figure 9b). As a result, the jump is absent in the blue curve of figure 8a but its "premonition" can be noticed in the rapid OV migration between $z = 10$ cm and $z = 15$ cm.

Noteworthy, in the previously considered examples of the edge diffraction [31], the "retrograde" segments of vortex lines and distances between the planes $z_1$ and $z_2$ were relatively short (between the critical points, a vortex line was nearly orthogonal to the propagation axis). This enabled to treat the jump as a single event (see section 3.2). Now, among the presented examples of the phase-step diffraction, only situations described by the red curves in figures 3a ($\varphi = 2\pi/3$, $a \approx -0.96$, $z = 0.56 z_R$) and 8a, b ($\varphi = 2\pi/3$, $a = -0.8$, $z \approx 0.3 z_R$) and the green curve in figure 6 ($\varphi = 7\pi/8$, $a \approx -1.05$, $z = 1.5 z_R$) can be interpreted in the "jump" spirit. In these cases, the dashed lines uniting points F and E (red in figures 3a, 8a, green in figure 6) give a simplified but generally correct impression of what really happens. The examples supplied by the red curve of figure 6 ($\varphi = 7\pi/8$, $-0.65 > a > -0.96$, $z = 0.56 z_R$) and by the cyan curves of figures 8a, c ($\varphi = 5\pi/6$, $a = -0.8$, $0.15 z_R < z < 0.65 z_R$) testify that between the "start" F and "finish" E of the "jump", the singular skeleton experiences a long and complex transformation that cannot be characterized merely as the "rapid translation from F to E". Accordingly, the corresponding dashed lines between F and E possess just a symbolic meaning and poorly characterize the real evolution of $V_1$. It should be noted that similar situations can also happen in the edge diffraction of OV beams but in the special circumstances of [28–31] they were rather exotic and could only be realized at very small propagation distances and/or under conditions corresponding to very small diffraction-induced OV displacements, thus being of minor practical interest. On the contrary, in the process of phase-step diffraction, the patterns of "extended jump" are rather typical and can be easily realized and practically employed.

This idea is supported by figure 9b which shows how the critical points (more exactly, "boundaries" of the Z-like meander $z_1$ and $z_2$) depend on the transverse phase-step position. It testifies that in case of strong perturbation ($\varphi = 5\pi/6$) and when the phase step is close to the beam axis ($|a| < 0.7$), the difference between $z_1$ and $z_2$ is high. With growing distance between the phase step and the incident beam axis, the meander gets shorter and ultimately vanishes: for the phase step with $\varphi = 5\pi/6$ the meander segment disappears at $z > 1.3 z_R$, for $\varphi = 2\pi/3$ at $z > 0.56 z_R$ and for $\varphi = \pi/2$ at $z > 0.2z_R$. In the far field, any 3D trajectory contains no Z-like irregularities, and the $a$-dependent trajectory of $V_1$ contains no jumps (cf. figure 5). Remarkably, figure 9b provides additional confirmation of the essential difference between the "weak" and "strong" phase-step-induced perturbations as classified in section 3.2. Behind the weakly-perturbing screen with $\varphi = \pi/2$, the Z-like segment of the 3D vortex line is relatively short (the solid and dashed blue lines almost coincide), and the simplified concept of "jump" [31] is perfectly applicable whereas in the case of strong perturbation, $\varphi = 5\pi/6$, only the detailed picture of the $V_1$ evolution with explicit involvement of the topological reactions of birth and annihilation is adequate (the cyan+black trajectory in figure 8a supplies a convincing example). Expectedly, the case of $\varphi = 2\pi/3$ demonstrates an intermediate situation where the "jump" concept sometimes works quite well (see the red curve figure 3a), sometimes is valid approximately (cf. the red+black line in figure 8a).

In case of positive $a$, the single critical point of the combined ($V_2 + V'$) 3D trajectory is of the main interest (figures 7 and 9a). In this case, no qualitative distinctions between the "strongly" and "weakly" perturbed behaviours can be detected. In all situations considered in this section, $V_2$ and $V'$ exist separately at the early stages of the diffracted beam propagation. If the phase step is far from the beam axis (right side of figure 9a), their trajectories converge at moderate post-screen distances ($z_1 \approx 0.2z_R - 0.65z_R$ at $a = 1.3$). When the phase step advances towards the beam axis (decreasing $a$), the critical point moves to the infinity, and the phase-step height $\varphi$ only affects the rate of this tendency. For $\varphi = 5\pi/6$ the $z_1$ growth is the steepest: initially (at $a = 1.3$), the critical point is the closest to the screen plane ($z_1 \approx 0.2z_R$) and approaches infinity when $a \approx 0.85$. If the phase-step height is $\varphi = 2\pi/3$, the corresponding red curve starts from $z \approx 0.45z_R$ at $a = 1.3$ and reaches the vertical asymptote near $a \approx 0.7$; the blue curve for $\varphi = \pi/2$ is "smoother" but, anyway, goes to infinity near $a \approx 0.6$. Generally, for every $\varphi$ there exists a range of $a$ values (more or less close to the incident beam axis) for which the separate $V_2$ and $V'$ exist through the whole propagation distance and survive in the far field (see, e.g., figure 5).

## 4. Conclusion

In this work, we have numerically investigated the localization and migration of OVs in the diffracted optical field obtained after a circular single-charged OV beam passes through the transparent screen with a rectilinear phase step (figure 1). The approach based on the Kirchhoff–Fresnel diffraction theory has shown interesting and promising possibilities for purposeful formation of optical fields with desirable and controllable singular structures and revealed additional ways for the physical characterization and diagnostics of the OV beams.

It is demonstrated that a transparent phase-step screen can be an efficient instrument for controlling the OV positions and the whole singular skeleton pattern in the diffracted field. The only input parameters are the phase step value $\varphi$ that varies within the interval $0 < \varphi < \pi$ and the phase step position $a$ with respect to the incident beam axis that is changed in the course of the transverse screen translation, $\infty > a > -\infty$, across the beam cross section. When the step is situated far from the incident beam center ($|a| \gtrsim 2$ in units of the beam waist radius), the only influence of the phase step is that, in the diffracted field cross section, the OV is slightly shifted from the nominal beam axis. With changing $a$, the OV evolves along the spiral trajectory, which expectedly unwraps, oppositely to the energy circulation in the incident OV, when the phase step moves towards the axis and wraps back when it moves further to the beam periphery. This spiral migration reflects the fundamental helical nature of the incident OV beam and can be used for studying its physical properties and spatial structure.

While the phase step is not very high ("weak perturbation", $\varphi \lesssim 2\pi/3$), the OV, in the diffracted beam cross section, describes a smooth closed loop upon the "full" screen translation, $\infty > a > -\infty$. With growing $\varphi$, the "strong perturbation" condition is realized so that the continuous evolution of the single OV is replaced by the complicated pattern including trajectories' discontinuities, jumps and topological reactions with the birth of "new" OVs, their apparently independent migration, and annihilation of the "old" ones. These features, observed in a fixed cross section, are associated with the complicated behavior of the 3D vortex lines: depending on the phase-step height $\varphi$ and position $a$, these lines may form Z-like fragments with the "backward" segments which manifest themselves as oppositely charged additional OVs. In contrast to the OV "jumps" observed in the edge-diffracted OV beams [29,31], these Z-like fragments may be extended to rather high propagation distances $z$, and, instead of strictly localized "jumps", the intricate multi-vortex arrays may stably exist up to the far field. An impressive difference from the edge-diffracted fields is that the phase-step diffraction provides many new possibilities of forming much more diversified, complicated and rich of details singular-skeleton structures – noteworthy, all this diversity originates from the simplest incident OV beam. It is equally important that these structures are obtained by rather simple means, and they can be deliberately modified and efficiently regulated via the change of parameters $a$, $\varphi$ and $z$.

The fundamental aspect of the results obtained in this paper is that they provide additional and pictorial demonstrations of the intrinsic helical nature and "hidden" rotational properties of the circular OV beams. Besides, based on the very simple and "palpable" examples, they reveal the fundamental topological features of singular optical fields, the singularities' evolution, their interactions, topological reactions, etc., which usually requires more complex means (see, e.g., [45]). Possible applications of the results may be associated with the OV diagnostics; for example, the general view and the direction of the spiral OV evolution at high $|a|$ can be used for detection of the incident OV sign. The sensitivity of the OV position to the phase-step parameters may be useful for controllable optical tweezers implementing desirable transportations of the trapped objects. The calculated OV trajectories presented in figures 3 – 5 supply a wide range of helpful facilities: from the very fine regulation available when the OV positions weakly depend on $a$ (for example, near $a = 0$ in curves of figures 3 and 4, and in all situations where $|a| \gtrsim 2$) to the very sensitive conditions near the jumps (see, e.g., red curve in figure 3a or green curve $V_1$ in figure 6). In the latter case, a tiny change in the phase-step position with respect to the incident beam axis may induce a considerable shift of the OV core in the diffracted field, which can be applied for the precise detection or measurement of small mechanical displacements and deformations. In particular, the results of the present paper can be used for sensitive detection and measurement of linear phase defects, e.g., for the surface microtopography characterization [13,38,39].

And, as a final remark, it should be mentioned that this paper represents only the first attempt towards the systematic study of the phase-step OV-beams' diffraction. There are many related problems that could not be properly considered within the limited frame of a single article. In particular, details of the 3D singular skeleton behaviour deserve a more scrupulous investigation involving wider ranges of the phase-step heights $\varphi$ and positions $a$; a lot of intriguing and potentially impactful features of the diffracted field are expected in situations where $\varphi$ approaches closer to the "degenerate" case $\varphi = \pi$. Additional opportunities and interesting aspects of the singular skeleton behaviour are expected from the use of other incident fields, e.g., arbitrary multicharged $LG_{pm}$ modes. We hope that further studies, both theoretical and experimental, will elucidate the remaining questions and bring additional useful results.

**Appendix**

Here we present some simple analytical expressions describing the OV positions in the vortex LG beam diffracted by a phase step. In this appendix, we consider LG beams with arbitrary topological charge (azimuthal index) $m$ but still require the radial LG mode index to be zero.

The Kirchhoff–Fresnel integral (4) can be recast as

$$u(x,y,z,a) = u^{LG}(x,y,z) + \left(e^{i\varphi} - 1\right)Q(x,y,z,a) \tag{A1}$$

where

$$u^{LG}(x,y,z) = \frac{k}{2\pi i z} \int_{-\infty}^{\infty} dy_a \int_{-\infty}^{\infty} dx_a \, u_a(x_a, y_a) \exp\left\{\frac{ik}{2z}\left[(x-x_a)^2 + (y-y_a)^2\right]\right\} \quad (A2)$$

is the complex amplitude of the unperturbed incident beam (as if it propagated without any screen) and

$$Q(x,y,z,a) = \frac{k}{2\pi i z} \int_{-\infty}^{\infty} dy_a \int_{a}^{\infty} dx_a \, u_a(x_a, y_a) \exp\left\{\frac{ik}{2z}\left[(x-x_a)^2 + (y-y_a)^2\right]\right\}. \quad (A3)$$

The screen-induced OV displacements can be found as zeros of function (A1). Under assumption of small perturbation of the incident beam, we suppose that the searched displacements are small compared to the beam waist size $b$, and then the near-axis approximation of $u^{LG}(x,y,z)$ is valid. In this case, according to equations (A9) and (19) of [29] and (8) of [31],

$$u^{LG}(x,y,z) \simeq \frac{b}{\sqrt{|m|!}}\left(-\frac{iz_R}{b}\right)^{|m|+1} B r^{|m|} \exp(im\phi) \quad (A4)$$

where $z_R$ is determined by (8) and $r$, $\phi$ are the polar coordinates in the observation plane ($x = r\cos\phi$, $y = r\sin\phi$). Further, for $a \gg 1$ (in units of $b$), an asymptotic representation of function (A3) can be derived in the form

$$Q(x,y,z,a) \simeq \frac{b}{\sqrt{|m|!}}\left(-\frac{iz_R}{b}\right)^{|m|+1} D a^{|m|-1} \exp\left(-\frac{a^2}{2b^2}\right) \exp\left[\frac{ik}{z}\left(\frac{a^2}{2} - ax\right)\right] \quad (A5)$$

(see equations (19) and (A8) of [29] and (9) of [31]). The coefficients $A$ and $B$ in (A4) and (A5) are expressed as

$$B = \frac{1}{(z - iz_R)^{|m|+1}}, \quad D = \sqrt{\frac{i}{2\pi}\frac{k}{z}}(-iz_R)^{-|m|-1}\left[k\left(\frac{1}{z} + \frac{i}{z_R}\right)\right]^{-3/2} \quad (A6)$$

(here the relations (20) of [29] are adapted to an LG beam with the waist at the screen plane $z = 0$). Finally, after substitution of (A4) – (A6) into (A1) and equating the result to zero, the polar coordinates of the OV core can be found as

$$r = \left\{\left|\frac{D(e^{i\varphi}-1)}{B}\right| a^{|m|-1} \exp\left(-\frac{a^2}{2b^2}\right)\right\}^{1/|m|}, \quad (A7)$$

$$\phi + M\cos\phi = \frac{1}{m}\left\{\arg\left[D(e^{i\varphi}-1)\right] - \arg B\right\} + \frac{ka^2}{2mz} + \frac{2N-1}{m}\pi, \quad N = 0, 1, \ldots |m|-1 \quad (A8)$$

where

$$M = \frac{kra}{mz}. \quad (A9)$$

(cf. equations (21) and (22) of [29] and (19), (20) of [31]). In view of the near-axis approximation (A4), equations (A7) – (A9) describe only the "central" OV $V_1$ (see section 3), and are valid approximately for the initial segments ($a > 1$) of the $V_1$-trajectories in figures 3a, 3b, 4a, 4b, 6 and 7. In particular, (A7) and (A8) describe the monotonous growth of $r$ and $\phi$ with increasing $z$ and decreasing $a$ (remember that in our case $m = -1$), which explains the spiral-like motion of the OV core when $|M| < 1$, and possible "jumps" of the OV trajectories, with emergence of an additional OV and its further annihilation when $|M| > 1$. According to [31], the jump can occur near positions where

$$\cos\phi = 0, \quad \frac{d}{d\phi}(M\cos\phi) < 0. \quad (A10)$$

In case of the opaque-screen diffraction, for which the described asymptotic model was first derived [29], condition of large negative $a$ was meaningless ($a \ll -1$ meant that the incident beam is almost completely stopped by the screen). Now, for the transparent screen with transmission (2) and (3), the asymptotic formulas for large negative $a$ also make sense. In this situation, it is suitable to represent the Kirchhoff-Fresnel integral (4) in the form

$$u(x,y,z,a) = \frac{k}{2\pi i z} e^{i\varphi} \left[ \int_{-\infty}^{\infty} dy_a \int_{-\infty}^{\infty} dx_a\, u_a(x_a, y_a) \exp\left\{ \frac{ik}{2z}\left[(x-x_a)^2 + (y-y_a)^2\right] \right\} \right.$$
$$\left. + \left(e^{-i\varphi}-1\right) \int_{-\infty}^{\infty} dy_a \int_{-\infty}^{|a|} dx_a\, u_a(x_a, y_a) \exp\left\{ \frac{ik}{2z}\left[(x-x_a)^2 + (y-y_a)^2\right] \right\} \right] \quad (A11)$$

whence one obtains

$$u(x,y,z,a) = e^{i\varphi}\left[ u^{LG}(x,y,z) + (-1)^m \left(e^{-i\varphi}-1\right) Q(-x,-y,z,|a|) \right] \quad (A12)$$

(the signs of the variables are reversed in the second term of (A11), with account that for the LG beams $u_a(-x_a,-y_a) = (-1)^m u_a(x_a, y_a)$). Corresponding analogs of (A7) – (A9) for $a < 0$ acquire the forms

$$r = \left\{ \left| \frac{D(e^{-i\varphi}-1)}{B} a^{|m|-1} \right| \exp\left(-\frac{a^2}{2b^2}\right) \right\}^{1/|m|}, \quad (A13)$$

$$\phi + M\cos\phi = \frac{1}{m}\left\{ \arg\left[D(e^{-i\varphi}-1)\right] - \arg B \right\} + \frac{ka^2}{2mz} + \frac{2N}{m}\pi - \frac{1+(-1)^m}{2m}\pi, \quad (A14)$$

(they differ from (A7) – (A9) by replacements $\varphi \to -\varphi$, $\cos\phi \to -\cos\phi$, $a \to |a|$, and the last summand of (A14) appears due to the multiplier $(-1)^m$ in (A12)). According to (A14), the current azimuthal coordinate of the OV core decreases with $a \to -\infty$ ($d\phi/d(-a) \propto -ka/m < 0$), which explains the clockwise direction of the spirals' wrapping in figures 3a, 3b and curves for $V_1$ migration in figure 6 at large negative $a$. The "jump conditions" (A10) remain the same but for negative $a$ and $m = -1$ they mean that the jump can occur in the upper half-plane, which corresponds to the "jumps" between points F and E on the red curves of figure 3a, 6 and 8a, green curve $V_1$ of figure 6 and the cyan curve of figure 8a. Actually, the asymptotic expressions (A7) – (A10) and (A13), (A14) reflect the main qualitative features of the "central" OV migration. However, in application to the examples considered in this paper, the numerical accuracy of the asymptotic model is not high because the condition $|a| \gg 1$ is never realized.

**Acknowledgements**

This work was supported, in part, by the Ministry of Education and Science of Ukraine, project 582/18 (State Registration #0118U000198).

**Figure captions**

**Figure 1**. Scheme of the OV transformation by the transparent screen with rectilinear phase step. The screen is orthogonal to the incident beam axis $z$ and is placed in the transverse plane $z = 0$; the boundary between the two screen parts introducing different phase shifts is parallel to axis $y$; the arrow shows the transverse energy circulation in the incident OV beam. After the transformation performed in the plane $z = 0$, the beam structure is registered (e.g., by the CCD camera) in the observation plane positioned at a distance $z$ behind the screen. Further explanations see in text.

**Figure 2**. (a, c, e) Intensity and (b, d, f) phase profiles of the diffracted $LG_{0,-1}$ beam at a distance $z = 0.56z_R$ (30 cm) behind the screen with the phase step (a, b) $\varphi = \pi/3$ ($A = 1$, $B = 0.5 + 0.866i$) and (c – f) $\varphi = \pi$ ($A = 1$, $B = -1$); insets in panels (a) and (c) show the phase surfaces in the near-axis area just behind the screen. The step position $a$ (see figure 1) is denoted by the vertical light-blue lines in panels (a, c, e) and equals to $a = 0$ in (a – d) and $a = 1$ in (e, f) (in units of $b$). Black contours in panels (b, d, f) indicate the intensity level 0.1 of the maximum, dashed ellipse in panel (f) shows the area where a pair of OVs is being formed (corresponds to the vicinity of the point B in figure 4a).

**Figure 3**. Migration of the OVs in the observation planes distanced from the screen by (a) $z = 0.56z_R$ (30 cm), (b) $z = 1.5z_R$ (80 cm) and (c) $z \to \infty$ (far field). Green, blue and red curves describe the cases $\varphi = \pi/3$, $\varphi = \pi/2$ and $\varphi = 2\pi/3$, correspondingly; markers indicate the current values of $a$ multiple of 0.1 in units of $b$, some values of $a$ are indicated explicitly near the markers; arrows show the direction of the OV motion when the screen translates in the negative $x$ direction ($a$ changes from $\infty$ to $-\infty$, see figure 1). Black line in panel (a) shows the trajectory of the "virtual" OV participating in the jump process shown by the dotted segment, asterisks mark the points of birth and annihilation (explanations in text); transverse coordinates are expressed in units of the current Gaussian envelope radius (10).

**Figure 4**. Migration of the OVs within the diffracted beam cross section when the transparent screen with the phase step $\varphi = \pi$ (see (2) and (3)) is translated in the negative $x$ direction ($a$ changes from $\infty$ to $-\infty$, see figure 1), for (a) $z = 0.56z_R$ (30 cm), (b) $z = 1.5z_R$ (80 cm) and (c) $z \to \infty$ (far field). Magenta curves show trajectories of the "main" near-axial OVs, black lines are the trajectories of the oppositely charged "accompanying" OVs participating in the topological reactions with the "main" ones. Arrows show directions of the OV migration; markers indicate the current values of $a$ multiple of 0.1 in units of $b$, some values of $a$ are indicated explicitly near the markers, asterisks show the points of topological reactions. Transverse coordinates are expressed in normalized units of (10).

**Figure 5**. OV trajectories in the far-field observation plane ($z \to \infty$) corresponding to the screen translation from $a = \infty$ to $a = -\infty$ (see figure 1) for the phase step values: (red curves) $\varphi = 2\pi/3$, (cyan curves) $\varphi = 5\pi/6$ and (magenta curves) $\varphi = \pi$. Asterisks denote the points of topological reactions, black lines are the trajectories of the oppositely charged "accompanying" OVs participating in the topological reactions, arrows show the directions of the OV migration, transverse coordinates are expressed in units of (10), which are equivalent to the far-field angular coordinates in units of the Gaussian envelope divergence $(kb)^{-1}$.

**Figure 6**. Trajectories of the "main" OVs $V_1$ (in the region $x/b_c < 0.3$) and their "partners" $V_2$ (at $x/b_c > 0.3$) in the observation planes distanced from the screen by (red) $z = 0.56z_R$ (30 cm), (green) $z = 1.5z_R$ (80 cm), (blue) $z = 3.7z_R$ (200 cm) and (brown) in the far-field ($z \to \infty$) for the phase step

height $\varphi = 7\pi/8$. Arrows show the OV motion when the phase-step screen translates from $a = \infty$ to $a = -\infty$ (see figure 1); the transverse coordinates are expressed in the normalized units of (10). Filled circles mark the trajectories' "vertices" (return points); empty circles denote some selected points (especially, starting points F of jumps (shown by dotted lines for red and green curves $V_1$, cf. figure 3a), asterisks show the points where the OVs emerge (are born) B, E and vanish (annihilate) A, D. Trajectories of the "accompanying" counterparts of $V_2$ ($V'$ and $V''$, see figures 4, 5) are not shown. Numbers near markers indicate corresponding values of the phase-step position $a$ (see figure 1). The jumps in the $V_1$ trajectories exist only at $z = 0.56z_R$ and $z = 1.5z_R$. It is visible how the trajectories of $V_1$ and $V_2$ gradually transform with growing $z$ to the symmetric far-field forms depicted in brown (cf. figures 3c, 4c and 5).

**Figure 7**. Trajectories of the "main" OVs $V_1$ (left and top scales), their "partners" $V_2$ and "accompanying" oppositely-charged $V'$ (right and bottom scales) observed behind the phase-step screen situated so that $a = 0.8$ (see figure 1) in the observation plane moving along the $z$-axis; markers show current distances in centimetres, the transverse coordinates are normalized by (10). Blue, red and cyan curves show the behaviour of $V_1$ and $V_2$ for $\varphi = \pi/2$, $\varphi = 2\pi/3$ and $\varphi = 5\pi/6$, correspondingly; black lines illustrate trajectories of the accompanying OV $V'$; asterisks denote the points where $V_2$ and $V'$ annihilate. The inset shows the 3D pattern of the vortex lines in the diffracted beam for the case $\varphi = 2\pi/3$; the longitudinal coordinate is normalized by (10). "Combined" trajectories $V_2$ and $V'$ form a continuous line, to which the transverse semitransparent plane $z_1$ is a tangent plane.

**Figure 8**. (a) Trajectories of the "main" OV $V_1$ observed behind the phase-step screen situated so that $a = -0.8$ (see figure 1) in the observation plane moving along the $z$-axis; markers show current distances $z$ from the screen plane in centimetres. Blue, red and cyan curves show the behaviour of $V_1$ for $\varphi = \pi/2$, $\varphi = 2\pi/3$ and $\varphi = 5\pi/6$, correspondingly; black lines illustrate trajectories of the "virtual" OVs. (b) and (c): 3D patterns of the OV evolution for $\varphi = 2\pi/3$ (red) and $\varphi = 5\pi/6$ (cyan); the "virtual" OV's trajectories (shown in black) form the retrograde segments of the continuous vortex lines (red + black and cyan + black trajectories of figure 8a are the transverse projections of these 3D lines). Points where the OV dipoles emerge (E) and disappear (D) are marked by asterisks, point F are the "starting points" of "jumps" (cf. figures 3a, 6); the transverse and longitudinal coordinates are normalized by (10). In the 3D patterns (b, c), the points E and D manifest as critical points where the 3D OV trajectories "turn back", and the transverse planes $z_1$ and $z_2$ are tangent to the 3D curves.

**Figure 9**. Longitudinal coordinates of the critical points of the 3D OV trajectories as functions of the phase-step position $a$ (see figure 1) for the phase step height $\varphi = \pi/2$ (blue) $\varphi = 2\pi/3$ (red) and $\varphi = 5\pi/6$ (cyan): (a) for the "combined" trajectory of $V_2$ and $V'$ (see the inset in figure 7) and (b) for the complex evolution of $V_1$ (see figures 8b, 8c). Solid (dashed) lines describe the behaviour of $z_1$ ($z_2$).

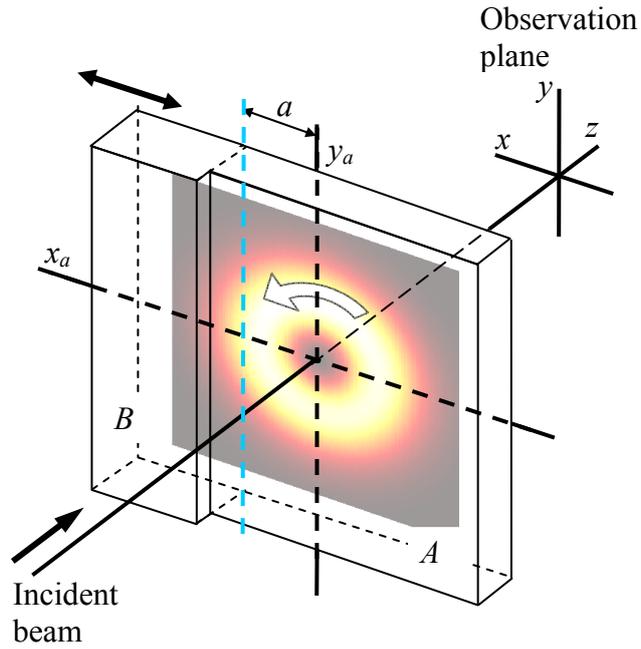

**Figure 1**. Scheme of the OV transformation by the transparent screen with rectilinear phase step. The screen is orthogonal to the incident beam axis $z$ and is placed in the transverse plane $z = 0$; the boundary between the two screen parts introducing different phase shifts is parallel to axis $y$; the arrow shows the transverse energy circulation in the incident OV beam. After the transformation performed in the plane $z = 0$, the beam structure is registered (e.g., by the CCD camera) in the observation plane positioned at a distance $z$ behind the screen. Further explanations see in the main text.

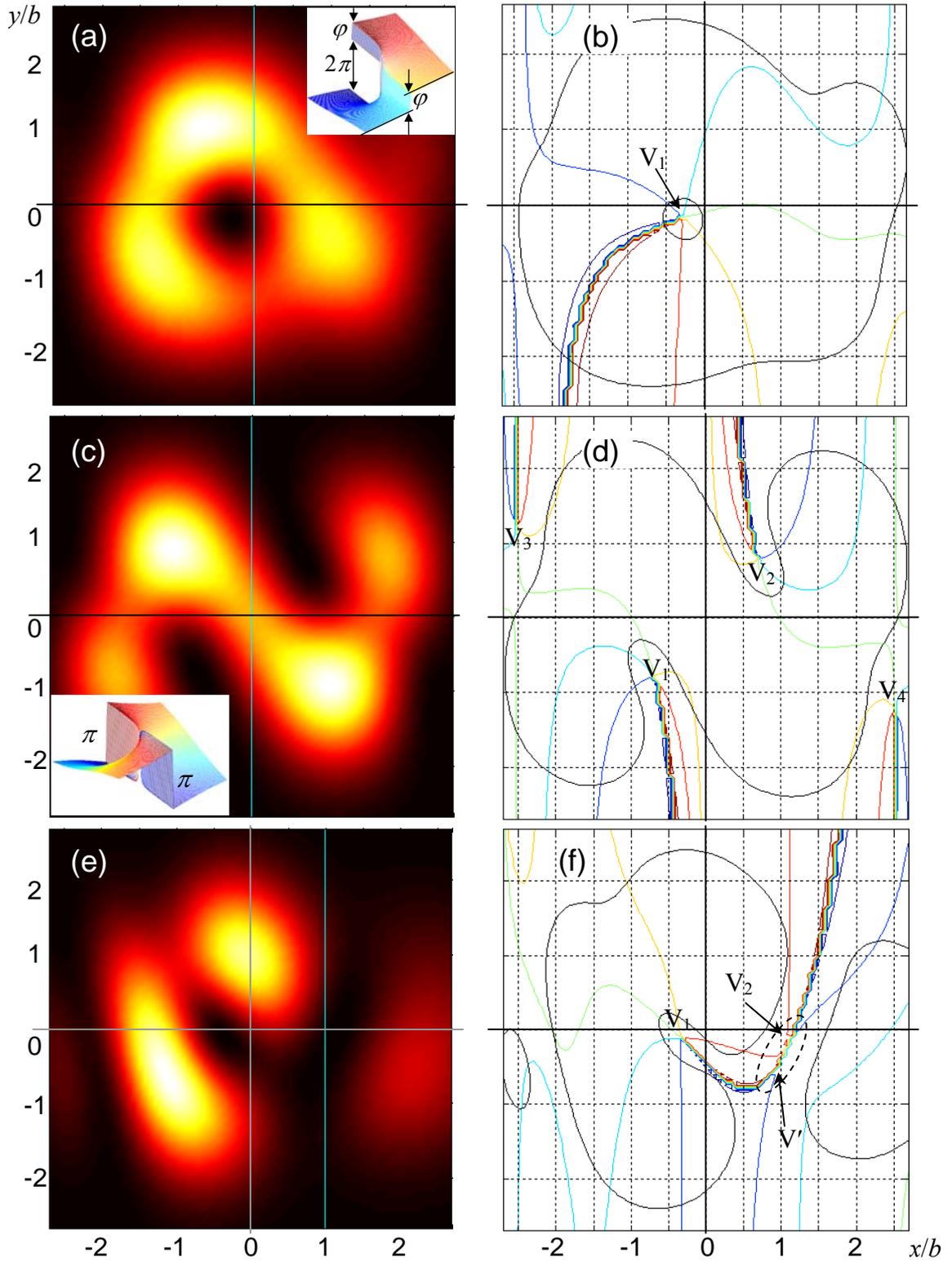

**Figure 2**. (a, c, e) Intensity and (b, d, f) phase profiles of the diffracted $LG_{0,-1}$ beam at a distance $z = 0.56 z_R$ (30 cm) behind the screen with the phase step (a, b) $\varphi = \pi/3$ ($A = 1$, $B = 0.5 + 0.866i$) and (c – f) $\varphi = \pi$ ($A = 1$, $B = -1$); insets in panels (a) and (c) show the phase surfaces in the near-axis area just behind the screen. The step position $a$ (see figure 1) is denoted by the vertical light-blue lines in panels (a, c, e) and equals to $a = 0$ in (a – d) and $a = 1$ in (e, f) (in units of $b$). Black contours in panels (b, d, f) indicate the intensity level 0.1 of the maximum, dashed ellipse in panel (f) shows the area where a pair of OVs is being formed (corresponds to the vicinity of the point B in figure 4a).

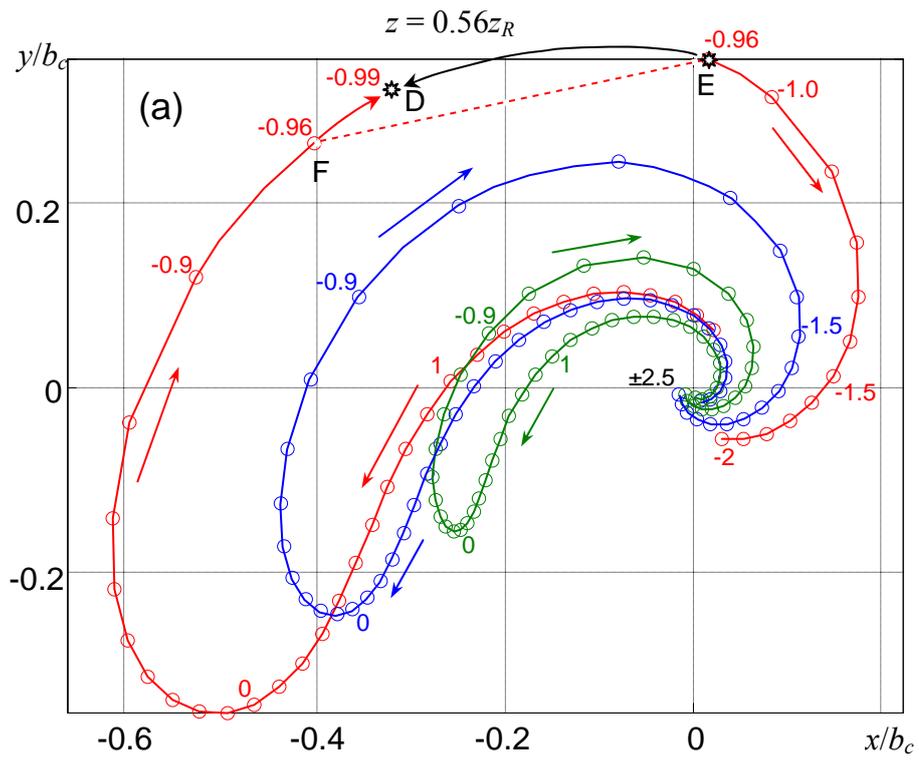

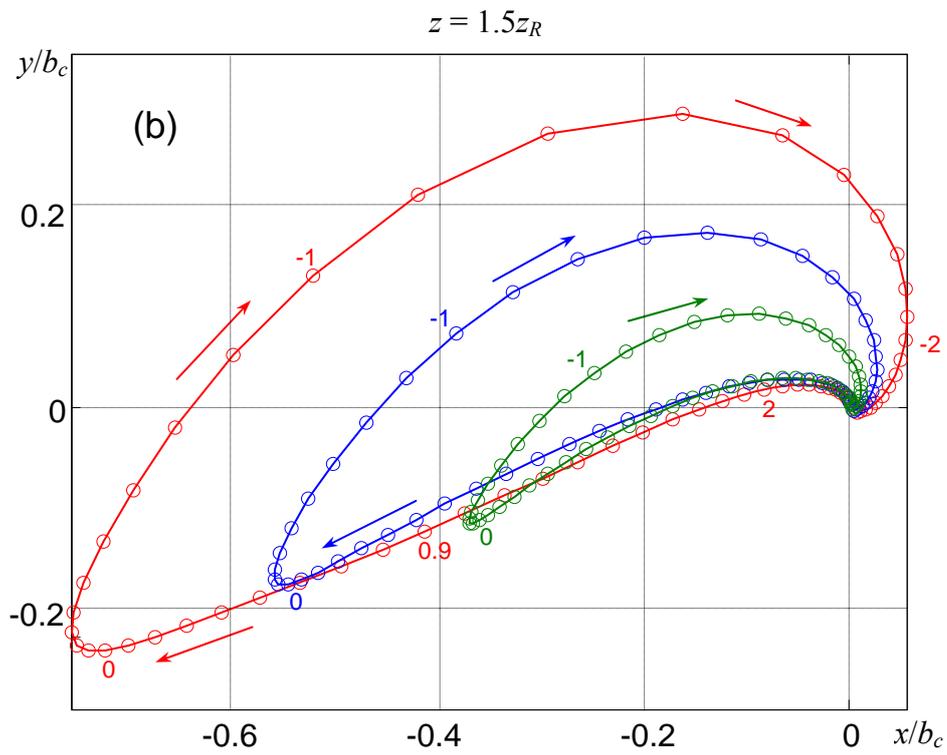

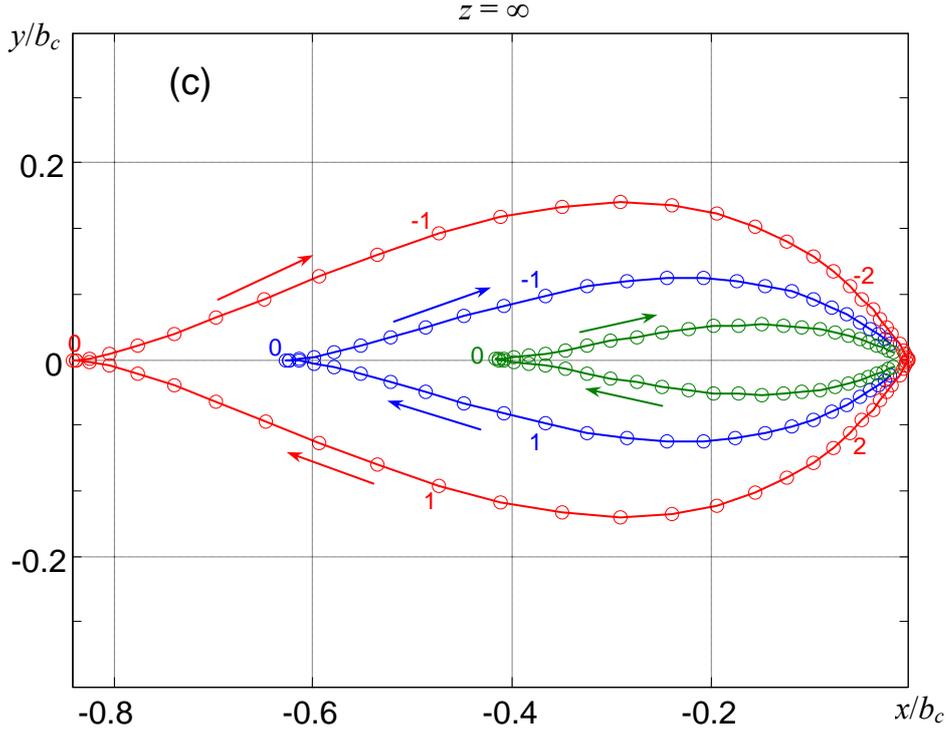

**Figure 3**. Migration of the OVs in the observation planes distanced from the screen by (a) $z = 0.56z_R$ (30 cm), (b) $z = 1.5z_R$ (80 cm) and (c) $z \to \infty$ (far field). Green, blue and red curves describe the cases $\varphi = \pi/3$, $\varphi = \pi/2$ and $\varphi = 2\pi/3$, correspondingly; markers indicate the current values of $a$ multiple of 0.1 in units of $b$, some values of $a$ are indicated explicitly near the markers; arrows show the direction of the OV motion when the screen translates in the negative $x$ direction ($a$ changes from $\infty$ to $-\infty$, see figure 1). Black line in panel (a) shows the trajectory of the "virtual" OV participating in the jump process shown by the dotted segment, asterisks mark the points of birth and annihilation (explanations in text); transverse coordinates are expressed in units of the current Gaussian envelope radius (10).

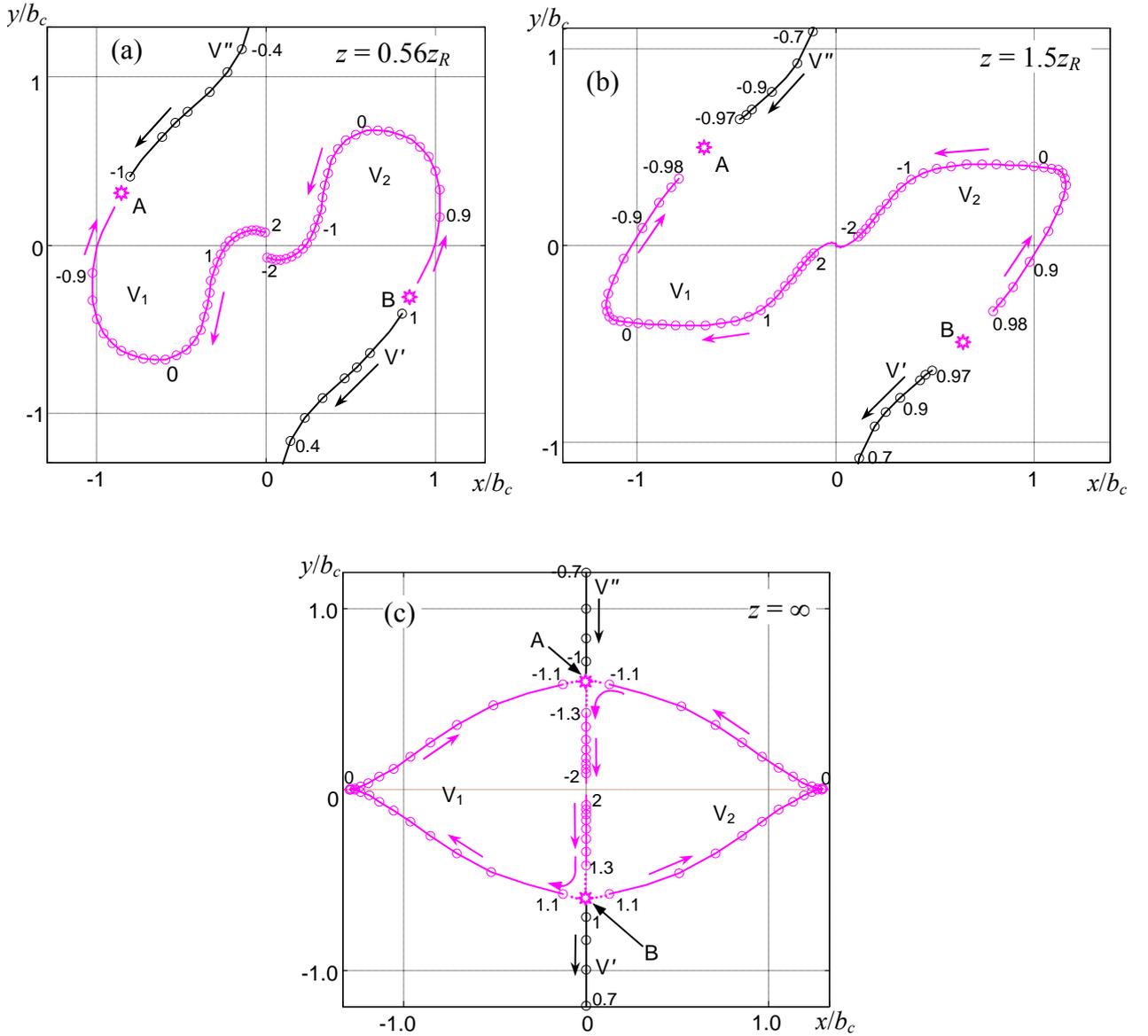

**Figure 4**. Migration of the OVs within the diffracted beam cross section when the transparent screen with the phase step $\varphi = \pi$ (see (2) and (3)) is translated in the negative $x$ direction ($a$ changes from $\infty$ to $-\infty$, see figure 1), for (a) $z = 0.56z_R$ (30 cm), (b) $z = 1.5z_R$ (80 cm) and (c) $z \to \infty$ (far field). Magenta curves show trajectories of the "main" near-axial OVs, black lines are the trajectories of the oppositely charged "accompanying" OVs participating in the topological reactions with the "main" ones. Arrows show directions of the OV migration; markers indicate the current values of $a$ multiple of 0.1 in units of $b$, some values of $a$ are indicated explicitly near the markers, asterisks show the points of topological reactions. Transverse coordinates are expressed in normalized units of (10).

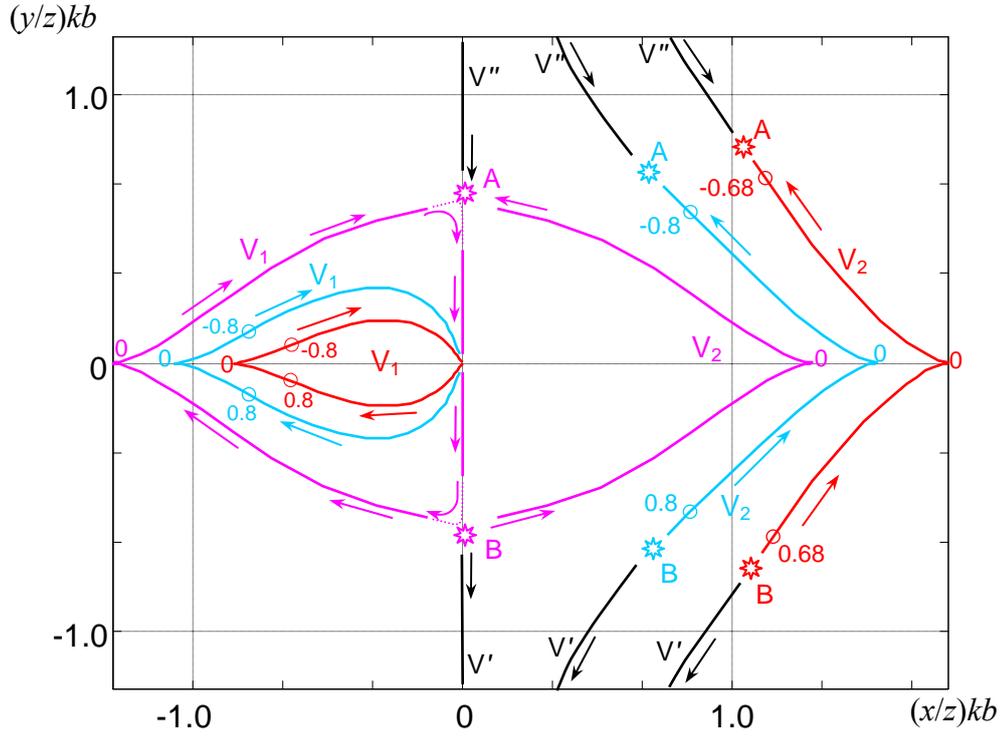

**Figure 5**. OV trajectories in the far-field observation plane ($z \to \infty$) corresponding to the screen translation from $a = \infty$ to $a = -\infty$ (see figure 1) for the phase step values: (red curves) $\varphi = 2\pi/3$, (cyan curves) $\varphi = 5\pi/6$ and (magenta curves) $\varphi = \pi$. Asterisks denote the points of topological reactions, black lines are the trajectories of the oppositely charged "accompanying" OVs participating in the topological reactions, arrows show the directions of the OV migration, transverse coordinates are expressed in units of (10), which are equivalent to the far-field angular coordinates in units of the Gaussian envelope divergence $(kb)^{-1}$.

**Figure 6**. Trajectories of the "main" OVs $V_1$ (in the region $x/b_c < 0.3$) and their "partners" $V_2$ (at $x/b_c > 0.3$) in the observation planes distanced from the screen by (red) $z = 0.56 z_R$ (30 cm), (green) $z = 1.5 z_R$ (80 cm), (blue) $z = 3.7 z_R$ (200 cm) and (brown) in the far-field ($z \to \infty$) for the phase step height $\varphi = 7\pi/8$. Arrows show the OV motion when the phase-step screen translates from $a = \infty$ to $a = -\infty$ (see figure 1); the transverse coordinates are expressed in the normalized units of (10). Filled circles mark the trajectories' "vertices" (return points); empty circles denote some selected points (especially, starting points F of jumps (shown by dotted lines for red and green curves $V_1$, cf. figure 3a), asterisks show the points where the OVs emerge (are born) B, E and vanish (annihilate) A, D. Trajectories of the "accompanying" counterparts of $V_2$ ($V'$ and $V''$, see figures 4, 5) are not shown. Numbers near markers indicate corresponding values of the phase-step position $a$ (see figure 1). The jumps in the $V_1$ trajectories exist only at $z = 0.56 z_R$ and $z = 1.5 z_R$. It is visible how the trajectories of $V_1$ and $V_2$ gradually transform with growing $z$ to the symmetric far-field forms depicted in brown (cf. figures 3c, 4c and 5).

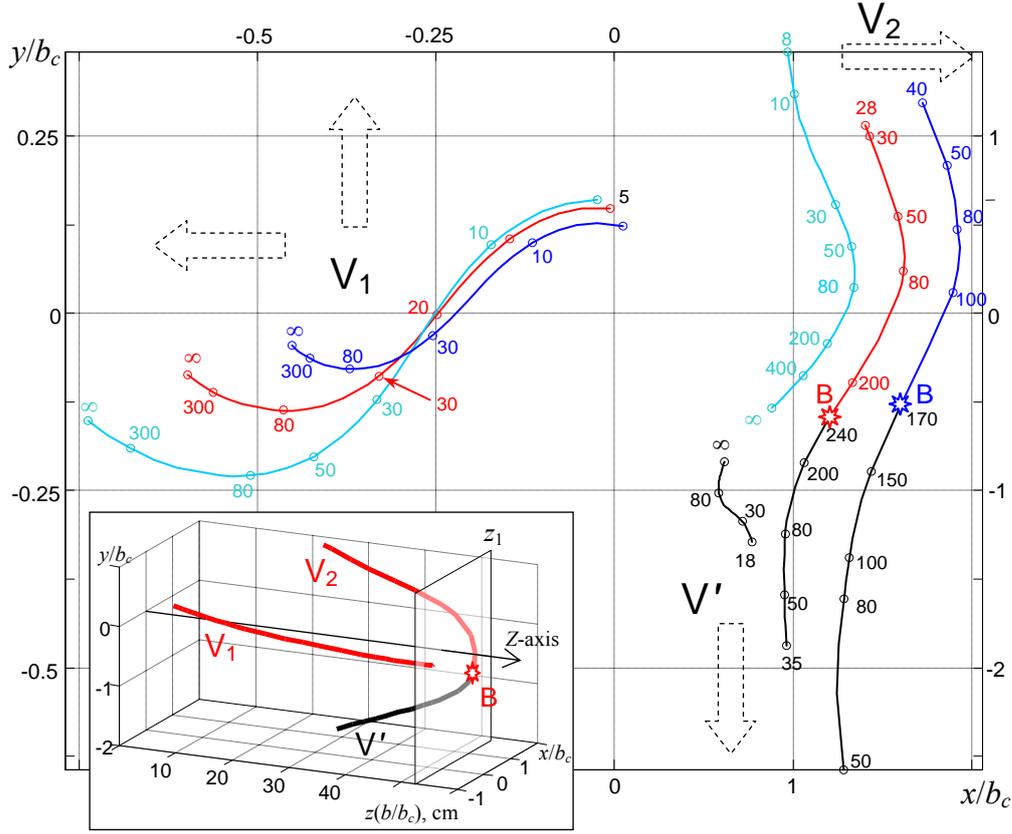

**Figure 7**. Trajectories of the "main" OVs $V_1$ (left and top scales), their "partners" $V_2$ and "accompanying" oppositely-charged $V'$ (right and bottom scales) observed behind the phase-step screen situated so that $a = 0.8$ (see figure 1) in the observation plane moving along the $z$-axis; markers show current distances in centimetres, the transverse coordinates are normalized by (10). Blue, red and cyan curves show the behaviour of $V_1$ and $V_2$ for $\varphi = \pi/2$, $\varphi = 2\pi/3$ and $\varphi = 5\pi/6$, correspondingly; black lines illustrate trajectories of the accompanying OV $V'$; asterisks denote the points where $V_2$ and $V'$ annihilate. The inset shows the 3D pattern of the vortex lines in the diffracted beam for the case $\varphi = 2\pi/3$; the longitudinal coordinate is normalized by (10). "Combined" trajectories $V_2$ and $V'$ form a continuous line, to which the transverse semitransparent plane $z_1$ is a tangent plane.

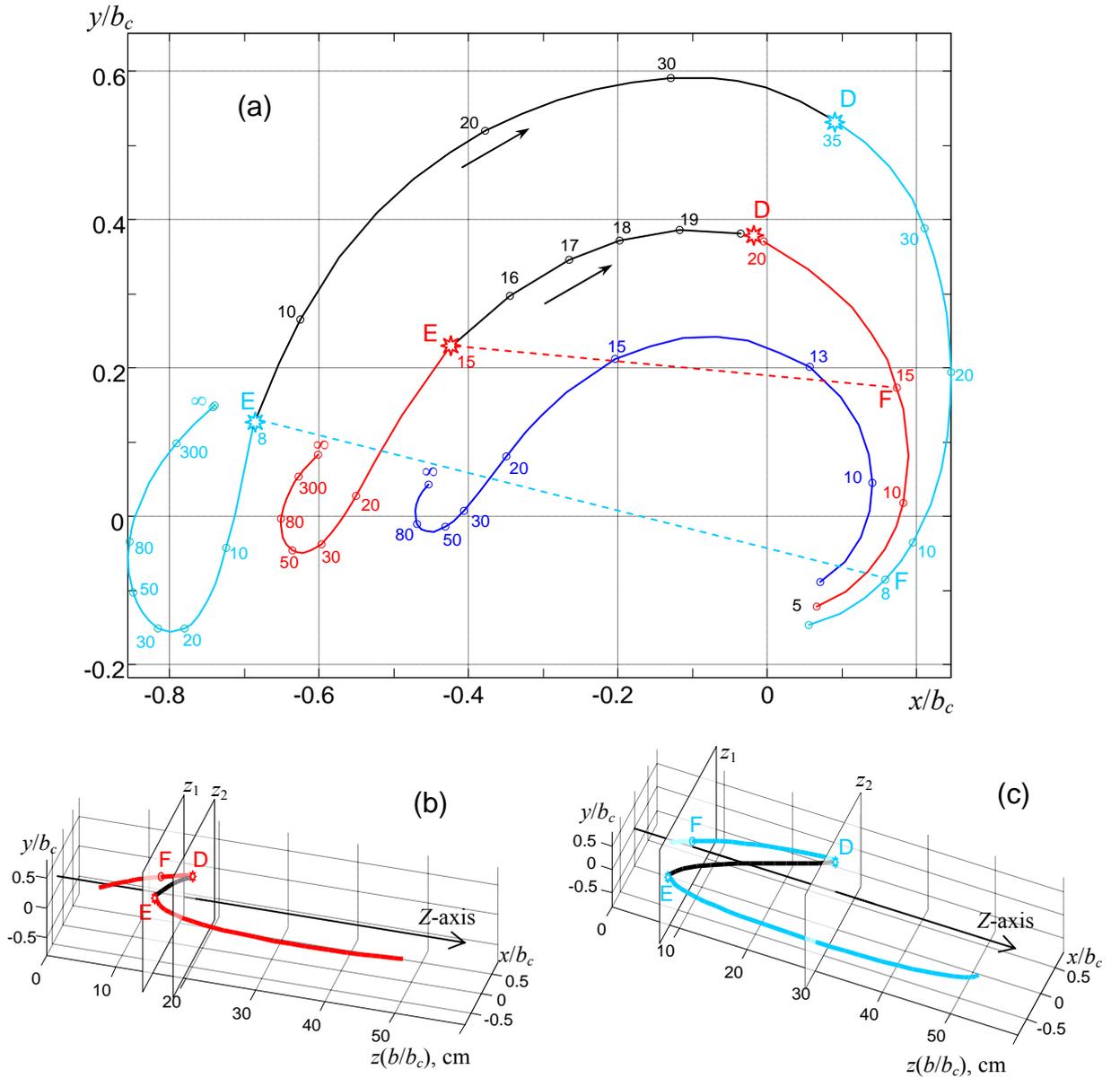

**Figure 8**. (a) Trajectories of the "main" OV $V_1$ observed behind the phase-step screen situated so that $a = -0.8$ (see figure 1) in the observation plane moving along the $z$-axis; markers show current distances $z$ from the screen plane in centimetres. Blue, red and cyan curves show the behaviour of $V_1$ for $\varphi = \pi/2$, $\varphi = 2\pi/3$ and $\varphi = 5\pi/6$, correspondingly; black lines illustrate trajectories of the "virtual" OVs. (b) and (c): 3D patterns of the OV evolution for $\varphi = 2\pi/3$ (red) and $\varphi = 5\pi/6$ (cyan); the "virtual" OV's trajectories (shown in black) form the retrograde segments of the continuous vortex lines (red + black and cyan + black trajectories of figure 8a are the transverse projections of these 3D lines). Points where the OV dipoles emerge (E) and disappear (D) are marked by asterisks, point F are the "starting points" of "jumps" (cf. figures 3a, 6); the transverse and longitudinal coordinates are normalized by (10). In the 3D patterns (b, c), the points E and D manifest as critical points where the 3D OV trajectories "turn back", and the transverse planes $z_1$ and $z_2$ are tangent to the 3D curves.

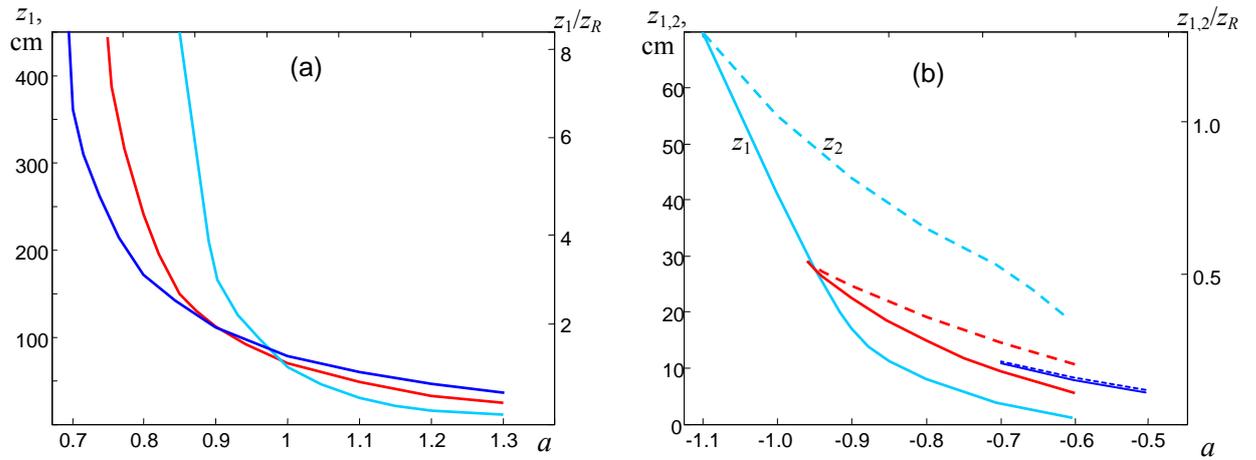

**Figure 9**. Longitudinal coordinates of the critical points of the 3D OV trajectories as functions of the phase-step position $a$ (see figure 1) for the phase step height $\varphi = \pi/2$ (blue) $\varphi = 2\pi/3$ (red) and $\varphi = 5\pi/6$ (cyan): (a) for the "combined" trajectory of $V_2$ and $V'$ (see the inset in figure 7) and (b) for the complex evolution of $V_1$ (see figures 8b, 8c). Solid (dashed) lines describe the behaviour of $z_1$ ($z_2$).